\documentstyle{article}

\begin{document}
\twocolumn[\hsize\textwidth\columnwidth\hsize\csname
@twocolumnfalse\endcsname
\date{ }
\title{The system with discrete interactions II:
Behavior of the relativistic particle in space and in time}
\author{M.Yudin}
\maketitle
\begin{center}
{\em
The Department of Mathematics and Statistics, 
The University of Melbourne, 
Parkville, Victoria,
Australia.}
\end{center}
\begin{abstract}
The principles of behavior of the system with discrete interactions are applied to description of motion of the relativistic particle. Applying the concept of non-local behavior both to position in space and to time, the apparently covariant equation of motion of the particle (Klein-Gordon) is defined. 
Based on the assumption that the form of the equation has to be invariant in inertial reference frames, the principle of constancy of the speed of light can be deduced. Consequently the basic statements of special relativity can be established in relation to the principles of discrete interaction.

The condition that Klein-Gordon equation was established without explicit use of the principle of correspondence gives a way to specify the area of applicability of the statement about the speed of light as the maximal speed of propagation of information. This would permit interpretation of superluminal phenomena with no contradictions to special relativity.

The conventional interpretation of time as a category which specifies succession and coexistence of events leads to interpreting of the Klein-Gordon and related equations (such as Maxwell or Dirac) as equations of the fields. This would justify the procedure of second quantization, so that the states of the field are identified with the sets of particles.  
\end{abstract}
\vskip2pc]

\large
\section{\rm Introduction}
The special theory of relativity in its classical formulation \cite{einstein} is based on the universally accepted principle of locality of interactions, the principle of relativity, that the character of behavior of a physical system is the same in inertial systems of reference, and constancy of speed of signal, conducted by light. This implies invariance of Maxwell equations, which were considered as fundamental and postulated independently. Alternatively, Maxwell equations are supposed to follow from local invariance of the Lagrangian under $U(1)$ transformations; Lagrangian of the system is essentially defined in a way that it has a covariant form. Similarly for other types of interactions, the system of principles used today, is the definition of the group of internal transformations, principles of relativity and constancy of the speed of signal (light).
The covariant form of equations of motion for particles (fields) essentially follows from these principles. The principle of constancy of the speed of signal applied to interactions of all types is interpreted as the principle of constant, maximal speed of propagation of an interaction. 

It may be noted, the statement about constancy of the speed of signal, conducted by light, may assume either constancy of the speed of signal as such (in the broad sense, a generic name for the mechanism of propagation of information between physical objects separated in space), or constancy of speed of possible propagators of the particular signals (such as photons), which are material objects with certain properties. The first assumption seems to be in contradiction with the well known superluminal phenomena in quantum systems: for a quantum system an interaction with one component of the system may have a seemingly instant effect on the other component. That means information about the state of this component is instantly reaching other components separated with the given one in space.  
The controversy of the predictions made by quantum theory and special relativity was noticed by Einstein, Podolky and Rosen in 1935 \cite{EPR}. This was attributed to the fact that quantum theory is not complete. The approach was further developed by Bell in 1964 \cite{bell}, with very many experiments to follow (to mention just a few \cite{weinfurter}, \cite{shih}). The theoretical conclusion of the thinking experiments discussed in the pioneering papers (as well as interpretation of the real experiments) was the concept of entangled state in the quantum system introduced by Schr\"odinger \cite{schrodinger}, which implies that measurement conducted on one component of the quantum system has an instant effect on the other component.
This may be interpreted as an indication of the phenomenon that the events inside the system have a common cause or, alternatively, information propagates within the system with an infinite speed. More strongly (the currently accepted point of view) the phenomena are interpreted in a way that the principle of locality is not applicable to the quantum system in entangled state (non-locality of quantum mechanics).
 
Note that the kind of correlation dependence between the components of the system postulated by the model of entanglement of states does not permit to pass information to the observer with an infinite speed and in this respect obviously does not contradict to special relativity. The model implies though, that cause-effect relation applies instantly to spatially separated components of the system, which is basically the definition of action-at-a-distance (albeit not dynamic).\footnote{The common argument that this is not a physical action, but statistical correlation must be scrutinized. Statistical correlation, as discussed by EPR, would imply {\em preexistence} of the particular values of the parameters, as prescribed to both components of the system: in disagreement with the principle of uncertainty.}
That means that the area of applicability of the principle of locality versus action-at-a-distance, which is the basic assumption of special relativity, should be clearly specified.

More precisely, if imply constancy of speed of propagators of the signal, we must specify due to which of their properties, these objects can move with the speed, which is constant in all inertial reference frames.\footnote{We do not mean here the obvious condition that the particles are massless, as the equation of motion of these particles initially was formulated in the covariant form.}
Note that generally speaking, the statement about constancy of speed of the signal, defined without a study of possible propagators (from which it may follow), in a sense is similar to postulating of elastic properties of solids based on the general thermodynamical approach, without considering the appropriate models of solid state physics.\footnote{A kind of argument used by Wheeler in speaking about the general relativity \cite{weeller}.} The theory is self-consistent but restricts the study of the physical reality at a certain level. 

It appears that the model of discrete interactions gives the way to obtain the results commonly associated with the postulate about constancy of the speed of light, without leading to the contradiction of the conventional approach. We consider this in more detail. 
We use the results and definitions established in \cite{yudin} without special references to it.
  
\section{\rm The principles of discrete interaction applied to behavior in time}
The principles of behavior of the system with discrete interactions (DI-system) as defined in \cite{yudin} can be formulated as follows 

(i) The parameters of a physical system are defined in process of its interaction;

(ii) Interaction is discrete.

The basic principles in particular imply that between interactions, which detect the particular value of the parameter, the DI-system has any possible value of this parameter, what makes a major difference with the behavior of the classical system. In particular, that means that the description of the particle in space is fundamentally non-local. As a state between interactions (intermediate state), we prescribe to the particle the set of positions it may enter, not one single position for each moment of time as used to be in the classical theory. So detection of the particle in space is not similar to registering of existing position, but something like choosing of the particular element (position) from the set.

In \cite{yudin} the behavior of the system was considered for the case when time was prescribed to the DI-system continuously. Each distribution $\Psi({\bf r})$ may be uniquely identified by the value of the parameter $t$, which has a meaning of time as defined in the reference frame, so that g-probability $\Psi({\bf r},t)$ describes the state of the system continuously in time. In other words, the model specifies the DI-behavior in space and classical in time. Space coordinates were interpreted as physical parameters: interaction is needed to detect the value of the coordinate; time was considered as a variable: it was {\em prescribed} to the system independent of any interactions.

The essential generalization of the approach is to consider the case when time is also interpreted as a physical parameter: {\em the interaction is needed to detect the value of time, which may be prescribed to the system.} 
According to our general approach, we assume that interaction of this type also has a discrete character. That means that the particular single value of time may be prescribed to the system, only if detected in the particular moment, as specified by the clock in the reference frame. For discrete interaction, at the particular moment of time the system may not interact with any detector located at any point in space; for each state between detection (as specified by distribution $\Psi({\bf r}))$\footnote{Time can be measured as an interval associated with a process: either for motion of a body relative to another body, or for change of internal state. In either case time may be measured only in relation to change of the state of the system (otherwise, for a stationary state, time is being prescribed to the system).
Consider a body in motion. We prescribe to the body different values of time according to changes of its position, or more generally, position in space and internal state. If the state of the system is characterized by a distribution $\Psi({\bf r})$ we prescribe different values of time $t$ to the different distributions.} a set of values $t$ may be prescribed to the system.

Consider the case of a scalar particle. 
In between interactions, the state of the particle in space is described by the distribution $\Psi({\bf r})$. In the non-relativistic model \cite{yudin}, the set of distributions $\Psi({\bf r})$ is defined for the interval $\Delta t$ between interactions, so that for each $t_{0} \in \Delta t$, a distribution  $\Psi = \Psi(t_{0}, {\bf r})$ is specified. The single value of time $t=t_{0}$ is associated with the particular distribution $\Psi({\bf r})$. In the present model a set of values $t$ is associated with each distribution $\Psi({\bf r})$, or alternatively, a set of functions $\Psi({\bf r})$ is defined for any single particular value of the parameter $t$.

In \cite{yudin} we have defined an intermediate state of the particle in space as a set of positions which is prescribed to the particle for the particular moment in time. 
Taking into account that within the present model, for each distribution $\Psi({\bf r})$, or alternatively for each position ${\bf r}$, the set of values of time is prescribed to the particle, we introduce intermediate state in time.
\\

{\em Definition} An intermediate sate of the particle in time is the set of values of time, which may be prescribed to the particle for the given particular position in space. 
 
\section{\rm Equation of motion of a particle}
We establish the equation of motion of a particle based of the following
\\

{\em Definition}[$\dagger$]: Entering of an intermediate state in space for the particular moment in time can be presented as entering of an intermediate state in time for the given position.
\\

This may be interpreted as the property of time. Alternatively we use much weaker condition: we postulate existence of the parameter $t\in R$ with the property [$\dagger$]. {\em This parameter can be identified with time as conventionally defined, provided we can show that it has the properties of time, that is, can be used to specify the sequence of events.}

Consider motion of the particle without internal states. The DI-particle is defined as a physical system with the property 
\\

{\em Definition}[$\ddagger$]: If the DI-particle enters intermediate state in space which includes a position ${\bf r}$, the g-probability $\phi({\bf r} + \delta {\bf r})$ that this state also includes a position ${\bf r}+\delta {\bf r}$, rapidly decreases with the growth of $\delta {\bf r}$. 
\\

We generalize this by adding the same constraint for the value $t$ (for convenience we will call the values of $t$ moments).

The g-probability that the particle enters an intermediate state, which includes the position ${\bf r}$ at the moment $t$ may be estimated in two different ways. We define $P_{T}({\bf r},t)$ as g-probability of entering of an intermediate state in time for the given position ${\bf r}$, ``in vicinity'' of the given moment $t$. That is, if the particle stays in the state, which includes the given position ${\bf r}$ at the moment $t_{1}$, the same state (no ordering specified) contains ${\bf r}$ also at the moment $t$, so that by definition 
\begin{equation}
P_{T}({\bf r},t)=\int_{-\infty}^{\infty}\Psi({\bf r},t_{1})\phi({\bf r},t,t_{1})dt_{1},
\label{p1}
\end{equation}
here $\Psi({\bf r},t_{1})$ is the g-probability that the particle stays in intermadiate state, which includes ${\bf r}$ at the moment $t_{1}=t+\delta t$, $\phi({\bf r},t,t_{1})$ is g-probability that if the state of the system defined by its position ${\bf r}$ includes  $t_{1}$, it also contains $t$. According to the definition
\[
\phi({\bf r},t, t_{1}) = \phi({\bf r},t_{1},t).
\]

We also define $P_{R}({\bf r},t)$, as g-probability of entering of an intermediate state in space for the given moment $t$, in vicinity of the given position ${\bf r}$. That is, if the particle stays in the state which includes the position ${\bf r}_{1}$ at the given moment $t$, this state also contains ${\bf r}$, so that by definition
\begin{equation}
P_{R}({\bf r},t)=\int_{V}\Psi({\bf r}_{1},t)
\theta(t,{\bf r},{\bf r} _{1})d{\bf r}_{1},
\end{equation}
here $\Psi({\bf r}_{1},t)$ is g-probability that the particle at the moment $t$ enters the intermadiate state which includes the position ${\bf r}_{1}={\bf r}+\delta {\bf r}$, 
$\theta(t,{\bf r},{\bf r}_{1})$ is the probability that if the state contains ${\bf r}_{1}$ at the moment $t$ it also includes ${\bf r}$ at the same moment $t$. According to the definition
\[
\theta(t,{\bf r}, {\bf r}_{1}) = \theta(t,{\bf r}_{1},{\bf r}).
\]

Consider the special case of homogeneous space and parameter $t$. Suppose existence of the reference frame $A$, which is defined for homogeneous space and parameter $t$: the characteristics of behaviour of the particle in this system do not explicitly depend on its position in space or the value of $t$. We call this reference frame inertial.
For the inertial reference frame $A$, 
$\theta(t,{\bf r},{\bf r}+\delta {\bf r})=\theta(\delta{\bf r})$ 
and $\phi({\bf r},t,t+\delta t)=\phi(\delta t)$
do not depend on ${\bf r}$ and $t$.

According to the definition $\phi(\delta t), \theta(\delta {\bf r})$ are even functions of their arguments. For isotropic space $\theta(\delta {\bf r})=\theta(\delta r)$, depends only on the distance between two positions.
Recalling that for the particle the functions $\theta(\delta r),\phi(\delta t)$  are rapidly decreasing with the growth of their arguments and using the expansion  
$\Psi(t_{1})\equiv \Psi(t)+\delta t \frac{\partial\Psi}{\partial t}+
\frac{\delta t^{2}}{2} \frac{\partial^{2}\Psi}{\partial t^{2}}+\dots$, 
within the accuracy of $\tau^{2}\equiv(\delta t)^{2}$, we have 
\begin{eqnarray}
P_{T}({\bf r},t) &\simeq& \Psi({\bf r},t)\int_{-\infty}^{\infty}\phi(\tau)d\tau
\nonumber\\
&\;&+
\frac{1}{2}\frac{\partial^{2} \Psi({\bf r},t)}{\partial
t^{2}}\int_{-\infty}^{\infty}\tau^{2}\phi(\tau)d\tau,
\label{p_1}
\end{eqnarray}
and similarly using expansion for $\Psi({\bf r}_{1})$ in vicinity of ${\bf r}$,
within the accuracy of $\rho^{4}\equiv(\delta r)^{4}$, we have
\begin{eqnarray}
P_{R}({\bf r},t) &\simeq& 4\pi \Bigl[\Psi({\bf r},t)\int_{0}^{\infty}\rho^{2}\theta(\rho)d\rho 
\nonumber\\
&\;&+
\Delta\Psi({\bf r},t)
\frac{1}{2}\int_{0}^{\infty}\rho^{4}\theta(\rho)d\rho\Bigr].
\label{p_2}
\end{eqnarray}
According to the definition [$\dagger$]
\begin{equation}
P_{T}({\bf r},t)\equiv P_{R}({\bf r},t).
\label{basic}
\end{equation}
Assumed ${\int_{-\infty}^{\infty}\delta^{2}\phi(\tau)d\tau}\neq 0$, (\ref{basic}) can be presented as
\begin{equation}
\frac{\partial^{2} \Psi}{\partial t^{2}}=
c^{2}\Delta\Psi-m^{2}c^{4}\Psi,
\label{kg}
\end{equation}
which has the form of Klein-Gordon equation for the scalar particle with the mass $m$, here 
\begin{equation}
c^{2}=4\pi
\frac{\int_{0}^{\infty}\rho^{4}\theta(\rho)d\rho}
{\int_{-\infty}^{\infty}\tau^{2}\phi(\tau)d\tau},
\label{c}
\end{equation}
and 
\begin{equation}
m^{2}c^{4}=2\frac{\int_{-\infty}^{\infty}\phi(\tau)d\tau-4\pi\int_{0}^{\infty}\rho^{2}\theta(\rho)d\rho}
{\int_{-\infty}^{\infty}\tau^{2}\phi(\tau)d\tau}.
\label{m}
\end{equation}
\\

{\em Non relativistic limit.}
\\

Note that in our previous discussion, even though we were using the term moment of time, we only introduced the parameter $t$ with the property [$\dagger$]. This parameter can be identified with time if we show that it may be used to specify coexistence and succession of the events (property conventionally associated with time). On the other hand in the non relativistic case when define
\\
Schr\"{o}dinger equation, we use the property of time: the value $t$ is used to mark subsequent events. 

Schr\"{o}dinger equation can be obtained from the Klein-Gordon equation by the ordinary  
\\
method for the case, when g-probability is complex and can be presented in the form
$\Psi=\Psi_{0}e^{-imc^{2}t}$,
so that $\frac{\partial}{\partial t}\Psi_{0}\ll mc^{2}$. In this case substitution of $\Psi$ into (\ref{kg}) ($\hbar\equiv 1$) will give
\[
i\frac{\partial}{\partial t}\Psi_{0}=-\frac{1}{2m}\Delta\Psi_{0}
+O\Bigl(\frac{1}{c^{2}}\Bigr).
\]
The fact that Schr\"{o}dinger equation, established using conventional interpretation of time, can be also obtained from Klein-Gordon equation can be interpreted as in indication that {\em the parameter $t$ formally defined according to} [$\dagger$] {\em does have the meaning of time as defined in the classical system.} That means we can consider the behavior of the DI particle in the reference frame as conventionally defined by the set of scales (rods) in space and clocks.
\\

Consider now the equation of motion of a particle in the other reference frame $A'$, which is moving relatively to $A$ with the constant speed $v$ in the direction $x$.
The reference frame $A'$ is described by the set of coordinates ${\bf r}'=(x',y',z')$ and time $t'$. The essential choice of the scales
for $x'$ and $t'$ should be determined by the condition that for $v\rightarrow 0$
\begin{eqnarray}
x'\rightarrow x+vt,\nonumber\\
t'\rightarrow t.
\label{limit}
\end{eqnarray} 
In this case, provided that $A'$ describes homogeneous space and time, $x',y',z',t'$ should be linearly related to $x,y,z,t$. Without loss of generality consider $y=y'$, $z=z'$,
\begin{eqnarray}
t'&=&\alpha_{11}t+\alpha_{12}x, \nonumber \\
x'&=&\alpha_{21}t+\alpha_{22}x,
\label{transformation}
\end{eqnarray}
$\alpha_{11}, \alpha_{21}, \alpha_{22}\neq 0$.
The lengths in space and an interval of time defined in $A'$ will be
\begin{eqnarray}
\delta x'&=&\alpha_{22}\delta x, \;\;
\delta y'=\delta y, \;\;
\delta z'=\delta z, \nonumber\\
\delta t'&=&\alpha_{11}\delta t.
\label{delta}
\end{eqnarray}
All arguments used to establish (\ref{kg}) for $A$ can be repeated for $A'$, so the equation of motion of the particle will have exactly the same form as (\ref{kg}), though the coefficients in the equation, which depend on the choice of scale for unit length in space and time, may be different. So we have for $\Psi({\bf r}',t')$, which specifies the g-probability of entering the states in $A'$,
\begin{eqnarray}
\frac{\partial^{2} \Psi({\bf r}',t')}{\partial t^{'2}} &=&
c_{x}^{2}\frac{\partial^{2} \Psi({\bf r}',t')}{\partial x^{'2}}+
c_{y}^{2}\frac{\partial^{2} \Psi({\bf r}',t')}{\partial y^{'2}}
\nonumber\\
&+&
c_{z}^{2}\frac{\partial^{2} \Psi({\bf r}',t')}{\partial z^{'2}}-
\mu^{2}c^{4}\Psi({\bf r}',t'),
\nonumber\\
\;
\label{kg11}
\end{eqnarray}
here $\Psi({\bf r}', t')$ is g-probability of entering of the position ${\bf r}'$ at the moment $t'$ as defined in $A'$
\begin{eqnarray}
c_{x}^{2}=\frac{\int_{-\infty}^{\infty}(\delta x')^{2}\theta'
d(\delta x') d(\delta y') d(\delta z')}
{\int_{-\infty}^{\infty}\tau^{'2}\phi'd\tau'},
\label{c_x}
\end{eqnarray}
similarly for $c_{y}$ and $c_{z}$,
\begin{equation}
\mu^{2}c^{4}=2\frac{\int_{-\infty}^{\infty}\phi'd\tau'
-\int_{V}\theta' d(\delta x')d(\delta y')d(\delta z')}
{\int_{-\infty}^{\infty}\tau^{'2}\phi'd\tau'}.
\end{equation}
The functions $\phi'$, $\theta'$ are defined  
as 
\\
$\phi'(\delta x', \delta y', \delta z')=
\phi(\alpha_{22}\delta x, \delta y, \delta z)$ and 
$\theta'(\delta t')=\theta(\alpha_{11}\delta t)$.
In the case when
\begin{eqnarray}
\alpha_{11}=1+\epsilon_{1},\;\; |\epsilon_{1}< 1|,\nonumber\\
\alpha_{22}=1+\epsilon_{2},\;\; |\epsilon_{2}< 1|,
\end{eqnarray}
recalling the results of Appendix we obtain that 
\begin{eqnarray}
\mu=m,\nonumber\\
c_{x}^{2}=c_{y}^{2}=c_{z}^{2}=c^{2}.
\end{eqnarray}
The ratio (\ref{c}) is a constant (of a particle), invariant in inertial reference frames, if the intervals $\delta x$, $\delta t$ are transformed according to the formulae (\ref{delta}).

On the other hand, the equation of motion (\ref{kg11}) may be derived from (\ref{kg}) by transformation of coordinates and time, so that we have
\begin{eqnarray}
(\alpha_{11}^{2}-c^{2}\alpha_{12}^{2})\frac{\partial^{2} 
\Psi({\bf r}',t')}{\partial t^{'2}}-
2(\alpha_{11}\alpha_{21}\;\;\;\;\;\;\;\;\;\;
\nonumber\\
- 
c^{2}\alpha_{22}\alpha_{12})\frac{\partial^{2}\Psi({\bf r}',t')}{\partial t^{'}\partial x^{'}}=
\nonumber\\
c^{2}\Bigl((c^{2}\alpha_{22}^{2}-\alpha_{21}^{2})\frac{\partial^{2} \Psi({\bf r}',t')}{\partial x^{'2}}+
\;\;\;\;\;\;\;\;\;\;\;\;\;\;\;\;
\nonumber\\
\frac{\Psi({\bf r}',t')}{\partial y^{'2}}+
\frac{\Psi({\bf r}',t')}{\partial z^{'2}}\Bigr)
-m^{2}c^{4}\Psi({\bf r}',t').
\nonumber\\
\;
\label{kg1}
\end{eqnarray}
Equations (\ref{kg11}), (\ref{kg1}) are equivalent if
\begin{eqnarray}
\alpha_{11}\alpha_{21}-c^{2}\alpha_{22}\alpha_{12}=0,\nonumber\\
\alpha_{11}^{2}-c^{2}\alpha_{12}^{2}=c^{2}\alpha_{22}^{2}-\alpha_{21}^{2}=1.
\label{identity}
\end{eqnarray}
This, together with the condition (\ref{limit}), uniquely identifies the values $\alpha_{ij}$ as coefficients of Lorenz transformations.
\begin{eqnarray}
\lefteqn{c\alpha_{21}=\alpha_{11}=\cosh\gamma,}
\nonumber\\
&
\alpha_{12}=c\alpha_{22}=\sinh\gamma;\;\;\gamma=v/c.
\label{lorenz}
\end{eqnarray}

Consider the other type of particle.
Repeating the previous arguments, we define the equation of motion of the particle in the reference frame $A$ 
\begin{equation}
\frac{\partial^{2} \Psi({\bf r},t)}{\partial t^{2}}=
c_{1}^{2}\Delta \Psi({\bf r},t)
-c_{1}^{4}m_{1}^{2}\Psi({\bf r},t),
\label{KG}
\end{equation}
the coefficients in the equation are defined by the expressions of the form (\ref{c}), (\ref{m}), with the functions $\phi_{1}$, $\theta_{1}$, which specify the behavior of the given particle.

Similarly for the reference frame $A'$, recalling the results of the Appendix, we obtain the similar equation (with the same coefficients)
\begin{equation}
\frac{\partial^{2} \Psi({\bf r}',t')}{\partial t^{'2}}=
c_{1}^{2}\Delta\Psi({\bf r}',t')-c_{1}^{4}m_{1}^{2}\Psi({\bf r}',t').
\label{kg02}
\end{equation}
The equations are inferred independently in each reference frame based on the definition [$\dagger$], which, we suppose, has the general character and holds in all reference frames. On the other hand (\ref{kg02}) may be obtained from the (\ref{KG}) by transformation of coordinates and time according to the formulae (\ref{transformation}) which links the definitions of the reference frames. The necessary condition for that is 
\begin{equation}
\alpha_{11}\alpha_{21}-c_{1}^{2}\alpha_{22}\alpha_{12}=0,
\end{equation}
$\alpha_{ij}\equiv \alpha_{ij}(c_{1})$.
Comparing this with (\ref{identity}) we conclude that $c_{1}=c$, which is a fundamental constant: describes motion of an arbitrary particle.   

Note that (\ref{kg}) can be considered as an operator equation
\begin{equation}
\hat{H}^{2}\Psi=(c^{2}\hat{p}_{j}^{2}+
c^{4}m^{2})\Psi,
\label{oper}
\end{equation}
here 
$\hat{H}=i\frac{\partial}{\partial t}$, and
$\hat{p}_{j}=-i\frac{\partial}{\partial x_{j}}$, 
$j=1,2,3$ (we use $\hbar\equiv1$).
The meaning of the operators is the same as in the non relativistic case, which was discussed in \cite{yudin}. The formula (\ref{oper}) essentially implies the equation for eigenvalues of operators
\begin{equation}
E^{2}=c^{4}m^{2}+c^{2}p^{2},
\label{ein}
\end{equation}
which has to be invariant in inertial systems of reference.
This in particular implies that the set $(E, {\bf p})$ has properties of a vector in relation to Lorenz transformations. 

As we see the concepts of special relativity are appearing when consider equation of motion of the DI-particle, provided the parameter $t$ attributed to the particle has the property $[\dagger ]$. As noted above this parameter can be identified with time (with the conventional properties of succession and coexistence of events) based on the condition that the Schr\"{o}dinger equation can be obtained from the Klein-Gordon equation.   

\section{\rm Multi-component particles}
We considered motion of a scalar particle, that is the case when the state of the particle is described by a single function $\Psi({\bf r}, t)$. Similarly we can define equation of motion for a multi-component particle. 

We discussed motion of the particle in space. The essential generalization of the approach is to consider non-coordinate states as well.
Consider equation of motion of the particle with internal state described by the variable $\nu$. In this case we define  
\begin{equation}
P_{T}({\bf r},\nu,t)=\int\Psi({\bf r},t_{1},\nu_{1})\phi_{1}({\bf r};t,\nu;t_{1},\nu_{1})dt_{1}d\nu_{1},
\label{pt_1}
\end{equation}
$\phi_{1}({\bf r}; t,\nu; t_{1},\nu_{1})$ is g-probability that if the state of the particle for the given position ${\bf r}$ includes time $t$ and the internal state $\nu$, it also includes the time ${t_{1}}$ and internal state $\nu_{1}$.
\begin{equation}
P_{R}({\bf r},\nu,t)=\int\Psi({\bf r}_{1},\nu_{1},t)\theta_{1}(t;{\bf r},\nu;{\bf r}_{1},\nu_{1})d{\bf r}_{1}d\nu_{1},
\label{pt_2}
\end{equation}
$\theta_{1}(t; {\bf r},\nu; {\bf r}_{1},\nu)$ is g-probability that if the state of the particle for the given moment $t$ includes the position ${\bf r}$ and the internal state $\nu$, it also includes the position ${\bf r}_{1}$ and internal state $\nu_{1}$.

Here two main cases may be distinguished:

(i) the variable $\nu$ is continuous; no special constraints on the type of the functions 
\\$\theta_{1}(t; {\bf r},\nu; {\bf r}_{1},\nu)$, $\phi_{1}({\bf r}; t,\nu; t_{1},\nu_{1})$ are imposed. The DI-system enters an intermediate state which is defined as a set of positions and internal states $\{\nu\}$. 

(ii) the variable $\nu$ is discrete; the functions $\theta_{1}(t; {\bf r},\nu; {\bf r}_{1},\nu_{1})$, $\phi_{1}({\bf r}; t,\nu; t_{1},\nu_{1})$ have the form
\[
\theta_{1}(t; {\bf r},\nu; {\bf r}_{1},\nu_{1})=
\theta_{r}(t; {\bf r}; {\bf r}_{1})
\delta_{\nu,\nu_{1}}+\delta({\bf r}-{\bf r}_{1})
h_{\nu,\nu_{1}},
\]
\[
\phi_{1}({\bf r}; t,\nu; t_{1},\nu_{1})=\phi_{t}({\bf r}; t; t_{1})\delta_{\nu,\nu_{1}}+\delta(t-t_{1})h_{\nu,\nu_{1}},
\]
$\delta()$ is Dirac $\delta$-function, $\delta_{i,j}$ is Kroneker $\delta$-function, $h_{i,j}$ is a symmetric matrix. 
Transition between internal states is allowed only in the same point in space and in time. 

Consider discrete set of internal states $\nu\in S$, so that the behavior of the DI-particle is described by the set of functions $\Psi_{\nu}({\bf r},t)$. Expansion of the integrals in (\ref{pt_1}) and (\ref{pt_2}) leads to the formulae similar to (\ref{kg}) for $\Psi_{\nu}$. The equation of motion of the particle in the given inertial reference frame $A$ will be given by the formulae of the type (\ref{kg}) for each component $\Psi_{\nu}$. Recalling that in the other inertial reference frame $A'$, the equations of motion (established independently) should have the same form we should assume that either the set of functions $\Psi_{\nu}$ is the set of scalars, or more generally, this set represents components of a tensor or a spinor.

Recalling that Klein-Gordon equation defined for the set $\Psi_{\nu}({\bf r},t)$ with the rules of transformation as for a spinor represents Dirac equation; Klein-Gordon equation defined for the set $\Psi_{\nu}({\bf r},t)$  with the rules of transformation  as for a massless vector represents Maxwell equations,\footnote{
{\em Maxwell equations}
\\

Consider Klein-Gordon equation for the components of the vector $A=(A_{0},{\bf A})$
\begin{equation}
\frac{\partial}{\partial x_{k}}\frac{\partial}{\partial x^{k}}A_{i}=0.
\label{kg_A}
\end{equation}
The property that $A_{i}$ is a vector means that inner product of $A_{i}$ with any other vector is a scalar,
or alternatively
\begin{equation}
\frac{\partial^{2} A_{i}}{\partial x_{i}\partial x^{k}}= -\frac{4\pi}{c}j_{k}, 
\label{vec_A}
\end{equation}
where $j_{k}=-\frac{c}{4\pi}\frac{\partial q}{\partial x_{k}}$ is a vector.
For a tensor $F_{ik}$ defined as 
\[
F_{ik}=\frac{\partial A_{k}}{\partial x^{i}}-\frac{\partial A_{i}}{\partial x^{k}},
\]
using (\ref{kg_A}) and (\ref{vec_A}), we obtain
\begin{equation}
\frac{\partial F_{ik}}{\partial x^{i}}= -\frac{4\pi}{c}j_{k};
\end{equation}
this, with the identity
\begin{equation}
e^{ijkl}\frac{\partial F_{ik}}{\partial x^{j}}=0,
\end{equation}
where $e^{ijkl}$ is a unit anti-symmetric tensor, constitute the set of Maxwell equations.
\\

{\em Dirac equation}
\\

Consider Klein-Gordon equation for the components of the spinor $\eta=(\eta_{1}, \eta_{2})$
\begin{equation}
\frac{\partial}{\partial x_{k}}\frac{\partial}{\partial x^{k}} \eta_{i}\equiv \hat{D}_{1}\hat{D}_{2}\eta_{i}= m^2\eta_{i},
\label{kg_spin}
\end{equation}
here 
\[
\hat{D}_{1,2}=(\frac{\partial}{\partial x_{0}} \pm\sigma_{k}\frac{\partial}{\partial x_{k}}),
\]
$\sigma_{k}$, $k=1,2,3$ are Pauli matrixes. 
Recalling the generic property, that for a spinor $\eta_{i}$ and 4-vector $v^{i}$, $i=1,2,3$, the product $(v^{0}+\sigma_{i}v^{i})\eta_{i}$ transforms as a spinor, we define a spinor 
\begin{equation}
\mu^{j}=\frac{1}{m} \hat{D}_{2}\eta_{i};
\label{Dirac}
\end{equation}
substitution into (\ref{kg_spin}) gives
\begin{equation}
\hat{D}_{1}\mu^{j}-m\eta_{i}=0.
\label{Dirac1}
\end{equation}
The formulae (\ref{Dirac}), (\ref{Dirac1}) is a set of Dirac equations for the bispinor $(\eta_{i},\mu^{\dot{j}})$
}
etc. we can conclude that as long as the Klein-Gordon equation is established, the other types of equations of motion may not need to be inferred from the first principles.
That is, the principles of behavior defined for DI-system, which allow us to establish Klein-Gordon equation for the scalar particle, in the case of the multi-component particle will lead to the appropriate equations of the relativistic quantum theory.

\section{\rm The particle with internal state}
Consider now equations (\ref{pt_1}), (\ref{pt_2}) for the case when intermediate state is defined as a set of positions and internal states $\nu$.

We use $\phi(\delta t,\delta \nu)=\phi_{1}(t,\nu; t_{1},\nu_{1})$, where $\delta t = t_{1}-t$,
$\delta\nu=\nu_{1}-\nu$
and $\theta(\delta{\bf r},\delta\nu)=\theta_{1}({\bf r},\nu; {\bf r_{1}},\nu)$, where
$\delta{\bf r}={\bf r}_{1}-{\bf r}$. According to the definition
\begin{equation}
\phi(\delta t,\delta \nu)=\phi(-\delta t,-\delta \nu),
\end{equation}
and
\begin{equation}
\theta(\delta{\bf r},\delta\nu)=\theta(-\delta{\bf r},-\delta\nu).
\end{equation}
That is, if $\theta$ is presented as a sum of symmetric and anti-symmetric terms 
\begin{equation}
\theta=\theta_{s} + \theta_{a},
\label{phi_sa}
\end{equation}
so that
\begin{eqnarray}
\theta_{s}(\delta {\bf r}, \delta \nu)=\theta_{s}(-\delta {\bf r}, \delta \nu), \nonumber\\
\theta_{s}(\delta {\bf r}, \delta \nu)=\theta_{s}(\delta {\bf r}, -\delta \nu), 
\label{phi_s}
\end{eqnarray}
and
\begin{eqnarray}
\theta_{a}(\delta {\bf r}, \delta \nu)=-\theta_{a}(-\delta {\bf r}, \delta \nu), \nonumber\\
\theta_{a}(\delta {\bf r}, \delta \nu)=-\theta_{a}(\delta {\bf r}, -\delta \nu), 
\label{phi_a}
\end{eqnarray}
$\theta_{s,a}(\delta {\bf r}, \delta\nu )\equiv\theta_{s,a}(\delta {\bf r}, \delta\nu, t, {\bf r})$ is also a function of $t,{\bf r}$. Similarly for 
$\phi\equiv\phi_{a}+\phi_{s}$.

According to the general approach, the expansion of $\Psi({\bf r}, \nu)$ in (\ref{basic}) gives
\begin{eqnarray}
\int_{\nu}\int_{-\infty}^{\infty}
\phi({\bf r},t,\nu;\delta\nu,\delta t)
\;\;\;\;\;\;\;\;\;\;\;\;\;\;\;\;\;\;\;\;\;\;\;\;\;\;\;\;\;\;\;\;\;
\nonumber\\
\times\Bigl(\Psi + \delta t 
\frac{\partial\Psi}{\partial t}+ 
\delta\nu\frac{\partial\Psi}{\partial \nu}
+ 
\frac{\delta t^{2}}{2}\frac{\partial^{2}\Psi}
{\partial t^{2}}+
\nonumber\\
\frac{\delta\nu^{2}}{2}\frac{\partial^{2}\Psi}{\partial \nu^{2}}+ 
\delta t\delta\nu\frac{\partial^{2}\Psi}
{\partial t\partial\nu}
+o(\delta t^{3})
\Bigr)d(\delta t)d(\delta\nu)= 
\nonumber\\
\int_{\nu}\int_{V}
\theta({\bf r},t,\nu;\delta\nu,\delta{\bf r}) 
\;\;\;\;\;\;\;\;\;\;\;\;\;\;\;\;\;\;\;\;\;\;\;\;\;\;\;\;\;\;\;\;\;\;
\nonumber\\
\times\Bigl(\Psi + \delta x_{k} 
\frac{\partial\Psi}{\partial x_{k}}+ 
\delta\nu\frac{\partial\Psi}{\partial \nu}
+
\frac{\delta x_{k}\delta x_{j}}{2}\frac{\partial^{2}\Psi}
{\partial x_{k}\partial x_{j}}+
\nonumber\\
+
\frac{\delta\nu^{2}}{2}\frac{\partial^{2}\Psi}{\partial \nu^{2}}
+
\delta x_{k}\delta\nu\frac{\partial^{2}\Psi}
{\partial x_{k}\partial\nu}
+o(\delta x_{k}^{3})
\Bigr)d(\delta{\bf r})d(\delta\nu).\nonumber\\
\label{Schr31}
\end{eqnarray}
Recalling (\ref{phi_sa}), (\ref{phi_s}) and (\ref{phi_a}), we obtain 
\begin{eqnarray}
Q_{1}\frac{\partial^{2}\Psi}{\partial t^{2}}+
Q_{2}\frac{\partial^{2}\Psi}{\partial t\partial \nu}=
\;\;\;\;\;\;\;\;\;\;\;\;\;\;\;\;\;\;\;\;\;\;\;\;\;\;\;\;
\nonumber\\
R_{0}\Psi+
R_{1}\Delta\Psi + 
{\bf R}_{2}\frac{\partial^{2}\Psi}{\partial x_{i}\partial \nu}+
R_{3}\frac{\partial^{2}\Psi}{\partial \nu^{2}},
\label{QR}
\end{eqnarray}
with the coefficients given by the formulae
\begin{eqnarray}
R_{0}&=&\int_{\nu}\int_{-\infty}^{\infty}
\phi_{s}({\bf r}, t, \delta \nu, \delta{\bf r})
d(\delta\nu)d(\delta t)
\nonumber\\
&-&
\int_{\nu}\int_V
\theta_{s}({\bf r}, t, \delta \nu, \delta{\bf r})
d(\delta\nu)d(\delta {\bf r}),
\nonumber\\
\;
\end{eqnarray}
\begin{equation}
R_{1}=\frac{1}{2}\int_{\nu}\int_{V}
\theta_{s}({\bf r}, t, \delta \nu, \delta t)\delta r^{2}
d(\delta\nu)d(\delta {\bf r}),
\end{equation}
\begin{equation}
Q_{1}=\frac{1}{2}\int_{\nu}\int_{-\infty}^{\infty}
\phi_{s}({\bf r}, t, \delta \nu, \delta t)\delta t^{2}
d(\delta\nu)d(\delta t),
\end{equation}
\begin{equation}
{\bf R}_{2}=\int_{\nu}\int_V\theta_{a}({\bf r}, t, \delta \nu, \delta{\bf r})\delta x_{i}\delta \nu
d(\delta\nu)d(\delta {\bf r}),
\end{equation}
which, according to the definition, is a vector in $E^{3}$,
\begin{equation}
Q_{2}=\int_{\nu}\int_{-\infty}^{\infty}\phi_{a}({\bf r}, t, \delta \nu, \delta t)\delta t \delta \nu
d(\delta\nu)d(\delta {\bf r}),
\end{equation}
\begin{eqnarray}
R_{3}&=&\frac{1}{2}\Bigl(\int_{\nu}
\int_V\theta_{s}({\bf r}, t, \delta \nu, \delta{\bf r})\delta\nu^{2}
d(\delta\nu)d(\delta {\bf r})\nonumber\\
&-&
\int_{\nu}\int_{-\infty}^{\infty}
\phi_{s}({\bf r}, t, \delta \nu, \delta{\bf r})\delta\nu^{2}
d(\delta\nu)d(\delta t)\Bigr).
\nonumber\\
\;
\end{eqnarray}
Assume $R_{j}$, $Q_{j}$ do not explicitly depend on ${\bf r}$, $t$, in this case, similarly to the previous, we use notations
\begin{eqnarray}
c^{2}=R_{1}/Q_{1}, \;\; c^{4}m^{2}=R_{0}/Q_{1}.\nonumber
\end{eqnarray}
Consider the case when
\begin{equation}
\Psi({\bf r}, \nu)=\exp(ie\nu)\psi({\bf r}),
\end{equation} 
which corresponds to the internal symmetry $U(1)$ as specified by the values of $\nu$,
with the constraint
\begin{equation}
({\bf R}_{2}/c)^2- Q_{2}^{2} =4R_{3};
\label{RQ_constr}
\end{equation}
after renaming of the variables
\begin{eqnarray}
c{\bf A}=-{\bf R}_{2}/2Q_{1}, \;\; A_{0}=Q_{2}/2Q_{1},\nonumber
\end{eqnarray}
we have
\begin{eqnarray}
\Bigl(\frac{\partial^{2}}{\partial t^{2}}
+
2ie A_{0}\frac{\partial}{\partial t}\Bigr)\Psi
&=&
c^{2}\Bigl(\Delta-
2i\frac{e}{c}{\bf A}\nabla\Bigr)\Psi
\nonumber\\
- c^{4}m^{2}\Psi
&-&e^{2}({\bf A}^{2}
-A_{0}^{2})\Psi.
\nonumber\\
\;
\label{ch_p}
\end{eqnarray}
For 
\begin{eqnarray}
\nabla{\bf A}=0, \;\;
\frac{\partial A_{0}}{\partial t}=0,
\end{eqnarray}
(\ref{ch_p}) can be presented as
\begin{equation}
\Bigl(i\frac{\partial}{\partial t}-eA_{0}\Bigr)^{2}\Psi=
c^{2}\Bigl(-i\frac{\partial}{\partial x_{i}}-\frac{e}{c}A_{i}\Bigr)^{2}\Psi+c^{4}m^{2}, 
\label{aaa}
\end{equation}
which has the form of equation of motion of a particle, with the charge $e$ ($e$ has discrete values, see \cite{yudin}) in the field $A({\bf r})$. The expression is obviously invariant in the inertial reference frames. It may be noted that the constraint (\ref{RQ_constr}), which implies relativistic invariance of the system, can be regarded as a necessary condition of existence of the system with $U(1)$ symmetry, as specified above.  
 
The vector field $A=A({\bf r})$ may be interpreted as an electromagnetic field: would we consider the equation of its motion in time (a standard equation of motion for massless vector DI-particle) we obtain Maxwell equations.

Note that the approach may be applied to the general case of internal symmetry with the resulting equation of the form (\ref{aaa}), which implies covariant derivative in the equation of motion.

\section{\rm Conventional interpretation of time: the concept of second quantization}
The definition [$\dagger$] and the Klein-Gordon equation was based on the statement that a set of values $t$ are to be identified with the particular distribution $\Psi({\bf r})$. This can be also stated as:
\\

{\em The set of distributions $\Psi({\bf r})$ is defined for the particle for each particular moment in time $t$.} 
\\

The statement is similar to definition of the DI-field as specified in \cite{yudin} in the environment where $t$ has conventional meaning of time, that is the variable which specifies coexistence and succession of events. In the more convenient notations $\phi({\bf r},t)\equiv \Psi({\bf r},t)$, where
$\phi({\bf r},t)=\sum_{\bf k}a_{\bf k}(t)e^{i{\bf k}{\bf x}}$, can be interpreted as a DI-field with g-probability $\Psi(a_{\bf k})$, so that the function $\Psi(a_{\bf k})$ satisfies Schr\"{o}dinger equation in the vector space $\{a_{\bf k}\}$.

As demonstrated in much detail in \cite{yudin}, the model of behavior of the DI-field is identical to the appropriate model of quantum field theory. In other words the statement [$\dagger$] and consequently the Klein-Gordon equation would lead to the definition of the appropriate DI (quantum) field, provided conventional interpretation of the variable $t$. This would be similar to second quantization as defined in quantum field theory.

It may be noted that the definition of the DI-field, established in relation to the appropriate classical field equation, leads to the concept of Fock space and emission-absorption operators 
\\
$A_{\bf k}, A^{\dagger}_{\bf k}$, so that the states of the DI-field are identified by the set of consecutive integer energy levels. This however does not imply that that these states are to be represented by the sets of point-like particles, as interpreted in quantum field theory. The condition that the states of the DI-field represent sets of particles is a consequence of the statement that the Klein-Gordon equation was established as the equation for the particle: the condition [$\ddagger$] was used.
In this case the different energy states $E_{n}$ of the DI-fields may be identified with the sets of particles.
This would be applicable to all the fields derived from Klein-Gordon equation, such as Maxwell or Dirac, provided the constraint that
commutation relation specified for the operators $A_{\bf k}, A^{\dagger}_{\bf k}$ would imply the particular energy spectrum end consequently  statistics for the ensemble of particles, which corresponds to the given DI-field. 

\section{\rm Discussion: Superluminal effects in quantum systems}
Based on the previous discussion we can come up with the definition of the signal as the process of propagation of a massless particle, which will have the same constant speed $c$ in all inertial reference frames. The fundamental property that the speed of the signal is the same in inertial reference frames in this case would follow from the definition [$\dagger$]. Clearly if the signal is used for sinchronization of the clocks in the different reference frames, or measuring of the distances, the appropriate quantities measured in the different inertial reference frames wold relate to each other as specified by Lorenz transformations. Based on that, the conclusion can be made that as long as interaction between spatially separated parts of the system is conducted by sending of a particle (or the appropriate wave), the speed of propagation of interaction can not exceed $c$.

Indeed, the equation of motion for a scalar DI-particle has the form of Klein-Gordon equation, invariant in inertial reference frames. For multi-component DI-particles, with the different rules of Lorentz transformations for their components, the equations of motion, which would follow from Klein-Gordon (such as Dirac or 
\\
Maxwell equations, etc.) would be similar to conventionally defined in the relativistic quantum theory. 
The conventional interpretation of time as a category, which specifies succession and coexistence of events, leads to interpreting of the Klein-Gordon and related equations as equations of the fields, which would justify the procedure of second quantization, with the states of the field identified by the sets of particles. Within all interpretations, the equations, established in explicitly covariant form, would confirm the statement about $c$ being the maximal speed of propagation of interactions, with the principle of locality of interactions clearly applicable. 

On the other hand, as long as we do not discuss the question what plays the role of the carrier of information inside the system, thus do not set a restriction that information is only conducted by a particle or a field, we do not need to assume that the speed of propagation of information is finite. That is basically pre-relativistic concept of action-at-a-distance revisited. It may be noted that in \cite{einstein} the signal is defined as a mechanism of passing of information with the speed constant in the different reference frames. The definition was subsequently re-interpreted and acquired the present generic meaning of the mechanism of passing of information in space and in time. Would the signal be interpreted in its original meaning, the relativistic approach would not exclude the concept of action-at-a-distance. 
This avoids controversies normally associated with EPR or Bell paradoxes in quantum mechanics. As long as we assume that in the particular systems information may propagate with an infinite speed, interaction applied to one of the components of the system will instantly affect all the others. 

Consider the thinking experiment suggested by Einstein at the Fifth Solvay Congress (we refer to the description given by Jammer \cite{jammer}).
A quantum particle is passing through a small hole in the diaphragm so that it is diffracted and then impinges upon a hemispherical scintillation screen located behind the diaphragm. The conventional interpretation of the experiment would assume that the wave function, which describes the particle after passing the hole, will be defined as a spherical wave which propagates towards the screen. It was noted by Einstein that if the wave function represents not an ensemble but an individual particle, before the scintillation takes place, the particle must be considered to be potentially present at all points in the vicinity of the screen. When the particle is detected at the particular point of the screen there is a ``peculiar action-at-a-distance'' which prevents the particle to be detected at another position on the screen.
The statement would be in perfect agreement with the present approach. The definition of the DI-particle implies that before interaction the set of positions in vicinity of the screen will be prescribed to the particle. The effect of action-at-a-distance noted by Einstein in this model would be a synonym to interaction, which by definition would mean selecting of the given position from the set.     
 
The approach should be generalized to include the case of detecting of the particular component of the composite system, which may consist of several components, separated in space. 
Consider fission of a particle into two, this is the system which may consist of one (fission is not detected) or two components (fission is detected), which is the common setting of EPR-type experiment. 
To define that the system consists of the two particles the detector should interact with it: before interaction, we cannot identify whether the system consists of one or two particles. In quantum mechanical terms we are talking about superposition of states one-two components. Interaction, which is required to define the number of components of the system, is applied to the system as a whole and should affect the system as a whole, that is, each of the components instantly.
The next interaction is not conducted on the system as a whole, as the system is identified to consist of two particles. The system is transferred to the particular state, in which the components are identified, that means interaction is conducted on the particular component, so that two separate particles are considered explicitly.

\pagebreak
\onecolumn
\section{\rm Appendix}
Consider the transformation
\begin{eqnarray}
t_{1}=t(1+\epsilon), && \epsilon < 1.
\label{trt}
\end{eqnarray}
Using the expansion 
\begin{equation}
\phi(t_{1}) \equiv \phi(t(1+\epsilon)) =
\sum_{k=0}^{\infty}\frac{\epsilon^{k}}{k!}\frac{\partial^{k}\phi}{\partial\epsilon^{k}}=
\sum_{k=0}^{\infty}\frac{\epsilon^{k}t^{k}}{k!}\frac{\partial^{k}\phi}{\partial t^{k}}\Bigr|_{\epsilon=0},
\end{equation}
we obtain
\begin{eqnarray}
\lefteqn{\int_{-\infty}^{\infty}t_{1}^{n}\phi(t_{1})dt_{1}=} \mbox{}
\nonumber\\
& &
\int_{-\infty}^{\infty}
\Bigl[\sum_{k=0}^{\infty}\frac{\epsilon^{k}t^{k+n}}{k!}
\frac{\partial^{k}\phi(t)}{\partial t^{k}}\Bigr](1+\epsilon)^{n+1}dt=
(1+\epsilon)^{n+1}\Bigl[\sum_{k=0}^{\infty}(-)^{k}\frac{(k+n)!}{k!n!}\epsilon^{k}\Bigr]
\int_{-\infty}^{\infty}\phi(t)t^{n} dt =
\int_{-\infty}^{\infty}\phi(t)t^{n}dt;\nonumber\\
\end{eqnarray}
while doing integration by parts for $\phi^{(n)}$, we use that
$\phi^{(n)}(t)=0$ for $t\rightarrow \pm\infty$.
The formula for $n = 0,2$ confirms invariance of the integrals as discussed in the paper.  


\begin{thebibliography}{99}

\bibitem{einstein} Einstein, A. (1905) {\em Annalen der Physik} 
{\bf 17}, 891.

\bibitem{yudin} Yudin, M. 
The system with discrete intearctions I:
Some Comments about the Principles of Quantum Theory.
{\em arXiv.org: quant-ph/0202082 }.

\bibitem{EPR} Einstein, A., Podolsky, B., Rosen, N. (1935) {\em Phys Rev} {\bf 45}, 777. 

\bibitem{bell} Bell, J.S. (1964) {\em Physics} {\bf 1}, 195.

\bibitem{weinfurter} Weinfurter, H., {\em Europhys Lett} {\bf 25}, 559.

\bibitem{shih} Shih, Y.H. Sergienko, A.V. {\em Phys Lett A} {\bf 186}, 29. 

\bibitem{schrodinger} Schrodinger, W. (1935) {\em Naturwissenschaften} {\bf 23}

\bibitem{FPQ} Greenberg, D.M., Zelinger, A. (ed.), {\em Fundamental Problems in quantum theory: A conference held in honor of Professor John A. Wheeler}
(New York: The New York Academy of Sciences, 1995).

\bibitem{weeller} Wheeler, J.A., Pregeometry: Motivations and 
Prospects, in
{\em Quatum Theory and Gravitation, ed. by A.R.Marlow} (New York: Academic
Press, 1980) pp 1-11.

\bibitem{landaul} Landau, L.D., Lifshitz, E.M., {\em The 
Classical Theory of Fields} (Oxford: Pergamon, 1971).

\bibitem{jammer} Jammer, M., {\em The Philosophy of Quantum Mechanics} (New York: Wiley, 1974).
\end{thebibliography}
\end{document}